\documentclass[twoside,leqno,twocolumn]{article}

\usepackage[letterpaper]{geometry}

\usepackage{ltexpprt}
\usepackage{hyperref}

\usepackage{graphicx}
\usepackage{float}
\usepackage{subfig}
\usepackage{multirow}
\usepackage{enumitem}
\usepackage{amsmath,amssymb}
\usepackage[switch]{lineno}

\begin{document}
	
	\newcommand\relatedversion{}
	\title{\Large DualVAE: Dual Disentangled Variational AutoEncoder for Recommendation}
	\author{
		Zhiqiang Guo\thanks{School of Computer Science and Technology, Huazhong University of Science and Technology. 
			\{zhiqiangguo, jianjunli\}@hust.edu.cn}
		\and Guohui Li\thanks{School of Software Engineering, Huazhong University of Science and Technology. guohuili@hust.edu.cn.} \thanks{Corresponding author.}
		\and Jianjun Li$^*$
	    \and Chaoyang Wang\thanks{Wuhan Digital Engineering Institute. sunwardtree@outlook.com}
    	\and Si Shi\thanks{Guangdong Laboratory of Artificial Intelligence and Digital Economy (SZ). shisi@gml.ac.cn}}
	
	\date{}
	
	\maketitle
	
	% Copyright Statement
	% When submitting your final paper to a SIAM proceedings, it is requested that you include
	% the appropriate copyright in the footer of the paper.  The copyright added should be
	% consistent with the copyright selected on the copyright form submitted with the paper.
	% Please note that "20XX" should be changed to the year of the meeting.
	
	% Default Copyright Statement
	\fancyfoot[R]{\scriptsize{Copyright \textcopyright\ 2024 by SIAM\\
			Unauthorized reproduction of this article is prohibited}}
	
	% Depending on which copyright you agree to when you sign the copyright form, the copyright
	% can be changed to one of the following after commenting out the default copyright statement
	% above.
	
	%\fancyfoot[R]{\scriptsize{Copyright \textcopyright\ 20XX\\
	%Copyright for this paper is retained by authors}}
	
	%\fancyfoot[R]{\scriptsize{Copyright \textcopyright\ 20XX\\
	%Copyright retained by principal author's organization}}
	
	%\pagenumbering{arabic}
	%\setcounter{page}{1}%Leave this line commented out.
	
	\begin{abstract} \small\baselineskip=9pt
		Learning precise representations of users and items to fit observed interaction data is the fundamental task of collaborative filtering. Existing studies usually infer entangled representations to fit such interaction data, neglecting to model the diverse matching relationships between users and items behind their interactions, leading to limited performance and weak interpretability. To address this problem, we propose a \underline{Dual} Disentangled \underline{V}ariational \underline{A}uto\underline{E}ncoder (DualVAE) for collaborative recommendation, which combines disentangled representation learning with variational inference to facilitate the generation of implicit interaction data. Specifically, we first implement the disentangling concept by unifying an attention-aware dual disentanglement and disentangled variational autoencoder to infer the disentangled latent representations of users and items. Further, to encourage the correspondence and independence of disentangled representations of users and items, we design a neighborhood-enhanced representation constraint with a customized contrastive mechanism to improve the representation quality. Extensive experiments on three real-world benchmarks show that our proposed model significantly outperforms several recent state-of-the-art baselines. Further empirical experimental results also illustrate the interpretability of the disentangled representations learned by DualVAE. 
	\end{abstract}
	
	\section{Introduction}
	\label{sec:introduction}
	% !TEX spellcheck = en_US
% !TeX root = ../main.tex

\begin{figure}[t]
	\centering
	\includegraphics[width=0.49\textwidth]{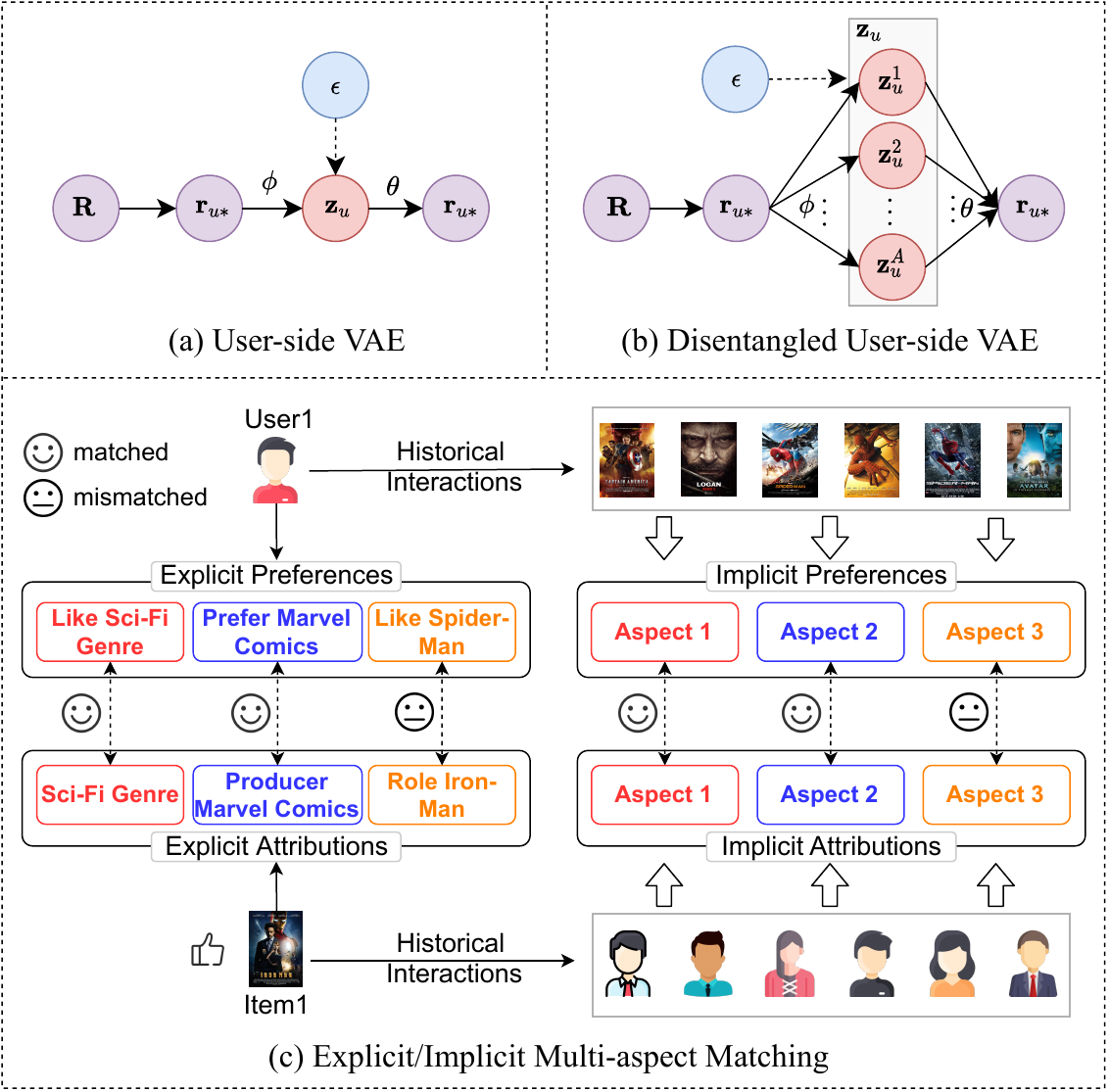}
	\caption{Illustration of (a) user-side VAE; (b) disentangled user-side VAE; and (c) explicit/implicit matching of multi-aspect features between users and items in a movie recommendation scenario. Words with different colors indicate different aspects.}
	\label{fig:motivation}
\end{figure}

The powerful ability of variational autoencoders (VAEs)~\cite{kingma2013auto} to account for uncertainty in the latent space has sparked a surge of interest in integrating variational inference into recommender systems~\cite{ yu2019vaegan, askari2021variational,ren2022variational}.
Traditional VAE-based CF methods~\cite{liang2018variational,truong2021bilateral,ma2022vae,cho2022stochastic} generally infer users' latent variables to reconstruct their interaction vectors, as shown in Figure~\ref{fig:motivation}~(a). These methods do not seek to learn deterministic user representations, but rather learn distributions over these representations, making them more suitable for sparse interaction data. However, due to the neglect of the complex entanglement of multiple potential factors behind user decisions, they are insufficient in learning robust and interpretable models. 

Recently, disentangled representation learning has received much attention in recommender systems, aiming at improving the representation quality by disentangling latent factors that govern user behaviors~\cite{ma2019learning,wang2020disentangled,zheng2021disentangling,zhao2022multi}. Most existing disentangled VAE-based CF methods~\cite{ma2019learning,zheng2021disentangling,wang2023causal} are typically designed based only on user-side interaction vectors to infer diverse user preference representations, while items are still entangled in the user vector space. As illustrated in Figure~\ref{fig:motivation}~(b), such a user-specific decoupling scheme usually ignores the diverse matching relationships between user preferences and item attributions, which makes the model performance still not satisfactory.

In realistic recommendation scenarios, items generally have multi-aspect attributions. Correspondingly, users have varying degrees of preference for different attributions of items. The preference of a user for an item can be regarded as the agglomeration of affinities between user preferences and item characteristics under different aspects. Take Figure~\ref{fig:motivation}~(c) as an example, the movie \emph{Item1} has multi-aspect attribution features, including \textit{Sci-Fi Genre}, \textit{Producer Marvel Comics} and \textit{Role Iron-Man}. \textit{User1}'s explicit preferences are reflected in multiple aspects of items, including \textit{Like Sci-Fi Genre}, \textit{Prefer Marvel Comics producer}, and \textit{Like Spider-Man}. In view of the high matching relationship between \textit{User1}'s preferences and \textit{Item1}'s attributions in multiple aspects, there is a high probability that \textit{Item1} is recommended to \textit{User1}. Predictably, coupled user and item representations cannot match such fine-grained correspondence to boost recommendation. Hence, decoupling and inferring multi-aspect features of items and users is necessary to enhance the modeling capability of VAE-based CF models, thereby improving their performance and interpretability. However, disentangled representation learning in such a scenario is non-trivial as it faces the following two challenges: 

\begin{itemize}[leftmargin=1.5em]
	% 如何解耦和推断
	\item \textbf{Firstly}, explicit user preference and item attribute information generally cannot be directly obtained in CF. We can infer implicit multi-aspect representations of users and items by only utilizing the available user-item interaction data, illustrated in Figure~\ref{fig:motivation}~(c). However, conventional binary interaction data typically reflect coupled interaction relationships between users and items. How to infer implicit decoupled multi-aspect representations of users and items based on observed binary interaction data becomes the main challenge.
	
	% 如何独立和一一对应
	\item \textbf{Secondly}, unconstrained inference of decoupled representations of users and items cannot guarantee independence among multi-aspect representations. Meanwhile, the learned aspect-level user preference may deviate from the corresponding aspect-level item representation, resulting in incorrect matching of multi-aspect representations between users and items. Hence, how to maintain independence among representations of different aspects for each user and item so that they do not deviate from the corresponding aspects is another tough challenge.
\end{itemize}

% 为解决上述问题，我们提出了双边解纠缠变分自编码器用于推荐。首先，为了解决第一个挑战，我们将解耦表征学习和双边变分自编码器相结合，利用基于原型的注意力机制解耦偏好数据，同时利用VAE推断用户和物品的多样性表征。值得注意的是，我们的模型同时对用户和物品进行推断，捕捉二元数据两边的不确定性，提高了其在稀疏偏好数据上的鲁棒性和性能。 此外，为了解决第二个挑战，我们提出解纠缠特征独立性建模，采用两级自监督对比学习来维护不同主题特征间的独立性和不同用户特征间的独立性。最后，在多个数据集上的实验结果表明了提出模型的有效性，杰出于对比方法。进一步的实验也表明模型生成的解耦表征的解释能力。
In this paper, we propose a novel \textbf{Dual} Disentangled \textbf{V}ariational \textbf{A}uto\textbf{E}ncoder (\textbf{DualVAE}) for collaborative recommendation with implicit feedback, which infers both user-side and item-side disentangled representations. Specifically, to address the first challenge, we transform the traditional VAE paradigm by unifying an attention-aware dual disentanglement module with a disentangled variational inference module and a joint generative module to infer multi-aspect latent representations for users and items. Further, we develop a neighborhood-enhanced representation constraint module to address the second challenge by employing self-supervised contrastive learning with neighborhood-based positive samples and two-level negative samples. Compared to standard VAE, DualVAE can capture multi-aspect uncertainty on both the user-side and the item-side, which helps improve its robustness and performance. Extensive experiments on three real-world benchmarks demonstrate the effectiveness of the proposed DualVAE model. Moreover, we also empirically show that the learned disentangled representations provide good support for explaining user behaviors.

	\section{Related work}
	\label{sec:relatedwork}
	% !TEX spellcheck = en_US
% !TeX root = ../main.tex
\subsection{Latent Representation Learning for CF}
To fit dyadic interaction data, much of the literature on collaborative filtering~\cite{1992UsingCF} focuses on latent representation learning.  Matrix factorization (MF)\cite{koren2009matrix} is a typical CF framework that directly embeds the IDs of users and items into latent feature space. Later, some studies~\cite{sedhain2015autorec,wu2016collaborative,he2017neural,zhu2019improving} model both users and items by introducing deep neural networks, which can be viewed as a nonlinear extension of MF. For instance, NeuMF~\cite{he2017neural} treats the recommendation problem with implicit feedback as a binary classification problem and fuses Generalized Matrix Factorization (GMF) with MLP. JCA~\cite{zhu2019improving} employs two separate autoencoders to simultaneously learn both user-user and item-item correlations. To further account for uncertainty in latent space, some researchers introduce VAEs~\cite{kingma2013auto} into collaborative filtering to improve model robustness~\cite{karamanolakis2018item,shenbin2020recvae,ren2022variational,gao2022mitigating,ma2022vae,cho2022stochastic}. For instance, MultiVAE~\cite{liang2018variational} extends VAEs to capture the latent variables of users with implicit feedback and estimate parameters via Bayesian inference. BiVAE~\cite{truong2021bilateral} infers the latent representations for users and items through bilateral VAEs. Despite their achievements, the entanglement of latent factors behind user behaviors, is mostly neglected by these methods, leading to weak interpretable results.

\subsection{Disentangled Representation Learning for CF}
Due to its robust performance and interpretability, disentangled representation learning~\cite{bengio2013representation} is gradually being introduced into VAE-based recommender systems. Most of these methods~\cite{ma2019learning,zheng2021disentangling,wang2023causal} usually decouple users' diverse preferences based on their collaborative interactions. For example, MacridVAE~\cite{ma2019learning} introduces categorical distributions to disentangle the macro-level and micro-level factors on different items via a variational autoencoder. Zheng \emph{et al.}~\cite{zheng2021disentangling} propose to structurally disentangle user interest and conformity by training with cause-specific data. Wang \emph{et al.}~\cite{wang2023causal} utilize structural causal models to generate causal representations that describe the causal relationship between latent factors. Nevertheless, these methods only unilaterally disentangle coarse-grained user preferences, while ignoring the relationship modeling between user preferences and item features in multiple aspects, and hence are not suitable for fitting binary interaction data.

	\section{Methodology}
	\label{sec:method}
	% !TEX spellcheck = en_US
% !TeX root = ../main.tex

%\begin{figure}[t]
%	\centering
%	\includegraphics[width=0.48\textwidth]{image/model2.pdf}
%	\caption{The overall framework of DualVAE with two aspects. Best viewed in color.}
%	\label{fig:model}
%\end{figure}

\subsection{Notations and Problem Formulation}
We consider an implicit collaborative recommendation that consists of a user set $\mathcal{U}$ with $m$ users and an item set $\mathcal{I}$ with $n$ items. The interaction data is denoted as $\mathbf{R} \in \mathbb{R}^{m \times n}$, where $r_{u,i} = 1$ is the observed interaction of user $u$ on item $i$. We use $\mathbf{r}_{u*} \in \mathbb{R}^{1 \times n}$ to denote $u$'s interaction vector corresponding to the $u$-th row in $\mathbf{R}$, and $\mathbf{r}_{*i} \in \mathbb{R}^{m \times 1}$ to denote the $i$-th column in $\mathbf{R}$ towards item  $i$. For traditional VAE-based CF models, the latent variables of per user and item are generally denoted by entangled $\mathbf{z}_u$ and $\mathbf{z}_i$, respectively. Considering that an item in general has multi-aspect features, we denote the item latent representation as $\mathbf{z}_i=[\mathbf{z}_i^{1}; \mathbf{z}_i^{2}; \dots ; \mathbf{z}_i^{A}] \in \mathbb{R}^{1 \times Ad}$, where $\mathbf{z}_i^{a} \in \mathbb{R}^{d}$ corresponds to $i$'s representation towards the $a$-th aspect, $A$ is the number of aspects, and $d$ is the embedding dimension. Similarly, user $u$'s latent representation is denoted as $\mathbf{z}_u=[\mathbf{z}_u^{1}; \mathbf{z}_u^{2}; \dots; \mathbf{z}_u^{A}]$, where $\mathbf{z}_u^{a} \in \mathbb{R}^{d}$ reflects $u$'s preference under the $a$-th aspect. Our objective in this work is to learn a robust VAE-based CF model to disentangle and infer multi-aspect latent representations of users and items from user-item interactions. After training, we can exploit the learned multi-aspect latent representations to perform top-$N$ recommendation.

\subsection{Overview}
Traditional VAE-based CF models generally define a user-side generative model that generates the observed data from the following distribution, 
\begin{equation}
   p_{\theta}(\mathbf{r}_{u*}) = \int p_{\theta}(\mathbf{r}_{u*}| \mathbf{z}_u) p(\mathbf{z}_u) ~ d\mathbf{z}_u,
\end{equation}
where  $p_{\theta}(\mathbf{r}_{u*}| \mathbf{z}_u) = \prod_{r_{u,i} \in \mathbf{r}_{u*}} p_{\theta}(r_{u,i}| \mathbf{z}_u)$ is a probability distribution over the $n$ items learned by a generative model with parameters $\theta$. Each observed interaction $r_{u,i} \in \mathbf{r}_{u*}$ is independently generated from the inferred user latent variable $\mathbf{z}_u$. $p(\mathbf{z}_u)$ is the prior of user variable $\mathbf{z}_u$. Different from the one-way generative paradigm, we propose a dual disentangled generative model to generate the observed interaction that encourages multi-aspect disentanglement of both users and items,
\begin{equation}
    p_{\theta}(r_{u,i}) = \int p_{\theta}(r_{u,i}| \mathbf{z}_u, \mathbf{z}_i, \mathbf{C}, \mathbf{P})p(\mathbf{z}_u)p(\mathbf{z}_i) ~ d\mathbf{z}_u\mathbf{z}_i,
\end{equation}
where the generation of $r_{u,i}$ is related to two latent variables $\mathbf{z}_u, \mathbf{z}_i$. $\mathbf{C}=\{\mathbf{c}_i\}^n_{i=1}$ and $\mathbf{P}=\{\mathbf{p}_m\}^m_{u=1}$ are the item-aspect and user-aspect probability matrices, in which $\mathbf{c}_i \in \mathbb{R}^{1 \times A}$ and $\mathbf{p}_u \in \mathbb{R}^{1 \times A}$ are two probability vectors drawn from two aspect distributions $p(\mathbf{c}_i)$ and $p(\mathbf{p}_u)$. $A$ is the number of aspects. We assume that $p(\mathbf{z}_u)=p(\mathbf{z}_u|\mathbf{C})$ and $p(\mathbf{z}_i)=p(\mathbf{z}_i|\mathbf{P})$, i.e., $\mathbf{z}_u$, $\mathbf{z}_i$, $\mathbf{C}$ and $\mathbf{P}$ can be independently generated. 

Our proposed DualVAE implements above distributions ($p(\mathbf{c}_i)$, $p(\mathbf{p}_u)$, $p(\mathbf{z}_u)$, $p(\mathbf{z}_i)$, and $p_{\theta}(r_{u,i}| \mathbf{z}_u, \mathbf{z}_i, \mathbf{C}, \mathbf{P})$) by four modules:
1) Attention-aware dual disentanglement (\textbf{ADD}) module that generates the aspect-aware probability distributions $p(\mathbf{p}_u)$, $p(\mathbf{c}_i)$ of users and items.
2) Disentangled variational inference (\textbf{DVI}) module that infers the posterior $p(\mathbf{z}_{1:|\mathcal{U}|}, \mathbf{z}_{1:|\mathcal{I}|} | \mathbf{R})$ over disentangled latent variables $\mathbf{z}_u$, $\mathbf{z}_i$ of users and items.
3) Joint generation (\textbf{JG}) module that utilizes the inferred user and item latent representations and aspect-aware probability matrices to reconstruct the observed interactions, thereby achieving $p_{\theta}(r_{u,i}| \mathbf{z}_u, \mathbf{z}_i, \mathbf{C}, \mathbf{P})$. 
4) Neighborhood-enhanced representation constraint (\textbf{NRC}) module that maintains the correspondence and independence between latent variables by introducing contrastive learning.
Next, we will detail the implementation of each module.

\subsection{Attention-aware Dual Disentanglement}
The ADD module is designed to generate the item-aspect and user-aspect probability vectors, which achieves the aspect assignment of users and items.  

\paragraph{item-side aspect disentanglement.} 
Considering that items usually have different degrees in different aspects, we assign an item-aspect probability matrix $\mathbf{C}=\{\mathbf{c}_i\}^n_{i=1}$ over all items in $\mathcal{I}$, where $\mathbf{c}_i$ is the aspect probability vector of item $i$. To generate the matrix $\mathbf{C}$, we assume $p(\mathbf{C})=\prod^n_{i=1} p(\mathbf{c}_i)$ and adopt the prototype-based attention mechanism to independently infer each aspect-level probability vector $\mathbf{c}_i$. Specifically, we introduce $A$ aspect prototypes $\{\mathbf{h}_a\}^A_{a=1}$ ($\mathbf{h}_a \in \mathbb{R}^d$) shared among items, and obtain the vector $\mathbf{c}_i$ from the item-aspect distribution $p(\mathbf{c}_i)$,
\begin{equation}
  	\mathbf{c}_i=[c^1_i; c^2_i; \dots; c^A_i]; ~\ 
  	c^a_i = \frac{exp(s(\mathbf{z}^a_i, \mathbf{h}_a))}{\sum_{a=1}^{A} exp(s(\mathbf{z}^a_i, \mathbf{h}_a))},
\end{equation}
where $\mathbf{z}^a_i$ is the $a$-th aspect-level latent representation of item $i$, $c^a_i \in \mathbb{R}_+$ is a probability that reflects the relation between item $i$  and aspect $a$, and $s(\cdot)$ is an affinity function (such as cosine) used to calculate the score of item $i$ under aspect $a$. 

\paragraph{user-side aspect disentanglement.}
Similarly, we also define $A$ preference prototypes $\{\mathbf{m}_a\}^A_{a=1}$ ($\mathbf{m}_a \in \mathbb{R}^d$) to calculate user-aspect probability vector $\mathbf{p}_u$ from the user-aspect distribution $p(\mathbf{p}_u)$,
\begin{equation}
   \mathbf{p}_u=[p^1_u; p^2_u; \dots; p^A_u]; ~\ 
   p^a_u = \frac{exp(s(\mathbf{z}^a_u, \mathbf{m}_a))}{\sum_{a=1}^{A} exp(s(\mathbf{z}^a_u, \mathbf{m}_a))},
\end{equation} 
where $\mathbf{z}^a_u$ is the $a$-th aspect-level latent representation of user $u$, and $p^a_u \in \mathbb{R}_+$ is a probability of user $u$ on the $a$-th preference. 

\subsection{Disentangled Variational Inference}
The DVI module is designed to infer the multi-aspect latent variables of users and items. In general, latent variables are drawn from prior distributions. We follow the convention to represent the prior over both user and item latent variables via the standard multivariate isotropic Gaussians, i.e., $p(\mathbf{z}_u)=\mathcal{N}(\mathbf{0}, \mathbf{I}), p(\mathbf{z}_i)=\mathcal{N}(\mathbf{0}, \mathbf{I})$, where $\mathbf{I}$ is a diagonal matrix. Given the user-item interaction matrix $\mathbf{R}$, the goal of variational inference model is to infer the posterior over the latent variables $p(\mathbf{z}_{1:|\mathcal{U}|}, \mathbf{z}_{1:|\mathcal{I}|} | \mathbf{R})$. Proverbially, the true posterior is intractable, and thereby precise inference is infeasible. We therefore exploit Variational Bayes~\cite{blei2017variational} to approximate this posterior distribution by a parameterized function $q_{\phi}(\mathbf{z}_{1:|\mathcal{U}|}, \mathbf{z}_{1:|\mathcal{I}|} | \mathbf{R}) = q_{\phi_u}(\mathbf{z}_{1:|\mathcal{U}|} | \mathbf{R})q_{\phi_i}(\mathbf{z}_{1:|\mathcal{I}|} | \mathbf{R})$. Obviously, the variational distribution firstly breaks the coupling between $\mathbf{z}_u$ and $\mathbf{z}_i$. Considering the statistical independence among users and items, we can set $q_{\phi_u}(\mathbf{z}_{1:|\mathcal{U}|} | \mathbf{R}) = \prod_{u}q_{\phi_u}(\mathbf{z}_{u} | \mathbf{r}_{u*})$ and $q_{\phi_i}(\mathbf{z}_{1:|\mathcal{I}|} | \mathbf{R}) = \prod_{i}q_{\phi_i}(\mathbf{z}_{i} | \mathbf{r}_{*i})$ to achieve the variational inference.

To further obtain disentangled multi-aspect latent representations of users and items, we assume that $q_{\phi_u}(\mathbf{z}_{u} | \mathbf{r}_{u*}) = q_{\phi_u}(\mathbf{z}_{u} | \mathbf{r}_{u*}, \mathbf{C}) = \prod^A_{a=1} q_{\phi_u}(\mathbf{z}^a_{u} | \mathbf{r}_{u*}, \mathbf{C})$ and $q_{\phi_i}(\mathbf{z}_{i} | \mathbf{r}_{*i}) = q_{\phi_i}(\mathbf{z}_{i} | \mathbf{r}_{*i}, \mathbf{P}) = \prod^A_{a=1} q_{\phi_i}(\mathbf{z}^a_{i} | \mathbf{r}_{*i}, \mathbf{P})$. Without loss of generality, we express the variational $q_{\phi_u}(\mathbf{z}^a_{u} | \mathbf{r}_{u*}, \mathbf{C})$ and $q_{\phi_i}(\mathbf{z}^a_{i} | \mathbf{r}_{*i}, \mathbf{P})$ as a multivariate
normal distribution with a diagonal covariance matrix,
\begin{equation}
 \begin{aligned}
   q_{\phi_u}(\mathbf{z}^a_{u} | \mathbf{r}_{u*}, \mathbf{C}) & \sim  \mathcal{N}(\boldsymbol{\mu}^a_u, \boldsymbol{\sigma}^a_u), \\
   q_{\phi_i}(\mathbf{z}^a_{i} | \mathbf{r}_{*i}, \mathbf{P}) & \sim 
   \mathcal{N}(\boldsymbol{\mu}^a_i, \boldsymbol{\sigma}^a_i),
 \end{aligned}
\end{equation} 
where $\boldsymbol{\mu}^a \in \mathbb{R}^d$ and $\boldsymbol{\sigma}^a \in \mathbb{R}^d$ are the mean and covariance of the disentangled variational distributions, parameterized by two shallow networks $f_{\phi_u}:\mathbb{R}^n \rightarrow \mathbb{R}^{2d}$ and $f_{\phi_i}:\mathbb{R}^m \rightarrow \mathbb{R}^{2d}$ with parameters $\phi_u$ and $\phi_i$,
\begin{equation}
  \begin{aligned}
	  \left[\boldsymbol{\mu}^a_u; \boldsymbol{\sigma}^a_u \right] &= f_{\phi_u}(\mathbf{r}^a_{u*}), ~ \mathbf{r}^a_{u*} = \mathbf{r}_{u*}^{\top} \otimes {\mathbf{c}^a},\\
	  \left[\boldsymbol{\mu}^a_i; \boldsymbol{\sigma}^a_i \right] &= f_{\phi_i}(\mathbf{r}^a_{*i}), ~ \mathbf{r}^a_{*i} = \mathbf{r}_{*i} \otimes \mathbf{p}^a,
  \end{aligned}
\end{equation}
where $\mathbf{c}^a \in \mathbb{R}^{n \times 1}$ and $\mathbf{p}^a \in \mathbb{R}^{m \times 1}$ are the $a$-th column of $\mathbf{C}$ and $\mathbf{P}$, respectively, representing the $a$-th aspect probability vectors over all items and users. $\otimes$ denotes element-wise product. $\mathbf{r}^a_{u*}$ and $\mathbf{r}^a_{*i}$ are calculated as the aspect-level interaction vectors of user $u$ and item $i$, respectively. It is clear to see that $\mathbf{r}_{u*}=\sum_a \mathbf{r}^a_{u*}$, and $\mathbf{r}_{*i}=\sum_a \mathbf{r}^a_{*i}$. Notably, under aspect $a$, the aspect-level probability vectors of items are employed to learn aspect-level latent representations of users, and vice versa. Such operations ensure that the disentangled representations of users and items are in a one-to-one correspondence at the aspect-level. Further, $\mathbf{z}^a_u$ and $\mathbf{z}^a_i$ are sampled by a \textit{reparameterization trick}~\cite{kingma2013auto,rezende2014stochastic},
\begin{equation}
 \mathbf{z}^a_u = \boldsymbol{\mu}^a_u + \boldsymbol{\sigma}^a_u \otimes \boldsymbol{\epsilon}, ~~~ \mathbf{z}^a_i = \boldsymbol{\mu}^a_i + \boldsymbol{\sigma}^a_i \otimes \boldsymbol{\epsilon}, 
\end{equation}
where $\boldsymbol{\epsilon}$ is a noise vector with $\boldsymbol{\epsilon} \sim \mathcal{N}(\mathbf{0}, \mathbf{I})$.

\subsection{Joint Generation}
Given the user and item latent representations $\mathbf{z}_u$ and $\mathbf{z}_i$ and the aspect probability matrices $\mathbf{C}$ and $\mathbf{P}$, the JG module is designed to predict the preference of a user towards an item by reconstructing their observed interactions.
Specifically, we generate the distribution $p_{\theta}(r_{u,i} | \mathbf{z}_u, \mathbf{z}_i)$ of user $u$'s preference towards item $i$ by a Poisson distribution~\cite{truong2021bilateral},
\begin{equation}
  p_{\theta}(r_{u,i} | \mathbf{z}_u, \mathbf{z}_i, \mathbf{C}, \mathbf{P}) = exp(r_{u,i} \rm log \it g_{\theta}(\mathbf{z}_u, \mathbf{z}_i)-g_{\theta}(\mathbf{z}_u, \mathbf{z}_i)),
\end{equation}
where $g_{\theta}(\mathbf{z}_{u}, \mathbf{z}_{i})$ is a differentiable function that utilizes disentangled latent representations of user $u$ and item $i$ to generate the joint observation $r_{u,i}$, 
\begin{equation}
\begin{aligned}
&g_{\theta}(\mathbf{z}_{u}, \mathbf{z}_{i}) =  \sum\nolimits_{a=1}^{A}  p^a_u \cdot c^a_i \cdot \sigma \left(\textsc{Skip}_{\theta}(\mathbf{z}^a_{u}, \mathbf{z}^a_{i})\right) \\
&\textsc{Skip}_{\theta}(\mathbf{z}^a_{u}, \mathbf{z}^a_{i}) = \mathbf{z}^a_u \odot \mathbf{z}^a_i + f_{\theta}(\mathbf{z}^a_{u}) \odot f_{\theta}(\mathbf{z}^a_{i})
\end{aligned}
\end{equation}
where $\textsc{Skip}_{\theta}(\cdot)$ is a skip-connection operation to avoid the issue of latent variable collapse~\cite{dieng2019avoiding}, $f_{\theta}(\cdot)$ is a non-linear function parameterized by $\theta$, $\sigma$ is sigmoid function, and $\odot$ denotes inner product. Notice $p_{\theta}(r_{u,i} | \mathbf{z}_u, \mathbf{z}_i, \mathbf{C}, \mathbf{P}) \propto g_{\theta}(\mathbf{z}_{u}, \mathbf{z}_{i})$ and $c^a_i$ and $p^a_u$ are regarded as aspect-level weights for item $i$ and user $u$.

To optimize our model parameters $\theta$, $\phi_u$ and $\phi_i$, we proceed with approximate inference and learning by maximizing the evidence lower bound (ELBO) $\sum_{u,i} \rm log \it p_{\theta}(r_{u,i})$. However, due to the sparsity of observed interactions, directly performing unbiased stochastic optimization on the above objective is inconvenient. We hence exploit the two-way nature of our model and perform alternate optimization in a Gauss-Seidel fashion~\cite{truong2021bilateral}. Specifically, we split the model parameters into user-related and item-related parts, and then alternately optimize them. Firstly, the user-side objective is updated with fixed item-related parameters,
\begin{equation}
	\begin{aligned}
		\mathcal{L}^u_{vae} &=  \mathbb{E}_{p(\mathbf{C})}[
		\mathbb{E}_{q_{\phi_u}(\mathbf{z}_u | \mathbf{r}_{u*}, \mathbf{C})}[\rm log \it p_{\theta_u}(\mathbf{r}_{u*} | \mathbf{z}_u, \mathbf{z}_{1:|\mathcal{I}|}, \mathbf{C})]  \\
		& - KL(q_{\phi_u}(\mathbf{z}_u | \mathbf{r}_{u*}, \mathbf{C}) \| p(\mathbf{z}_u))]
	\end{aligned}
	\label{ob_user}
\end{equation}
where $p_{\theta_u}(\mathbf{r}_{u*} | \mathbf{z}_u, \mathbf{z}_{1:|\mathcal{I}|}, \mathbf{C}) = \prod_i p_{\theta_u}(r_{u,i}| \mathbf{z}_u, \mathbf{z}_i, \mathbf{c}_i)$. The first term is interpreted as reconstruction error, while the second term defines the Kulback-Leibler divergence. Analogously, when keeping user-related parameters fixed, the item-side objective needs to be optimized,
\begin{equation}
\begin{aligned}
\mathcal{L}^i_{vae} &=  \mathbb{E}_{p(\mathbf{P})}[\mathbb{E}_{q_{\phi_i}(\mathbf{z}_i | \mathbf{r}_{*i}, \mathbf{P})}[\rm log \it p_{\theta_i}(\mathbf{r}_{*i} | \mathbf{z}_{1:|\mathcal{U}|}, \mathbf{z}_i, \mathbf{P})]  \\
& - KL(q_{\phi_i}(\mathbf{z}_i | \mathbf{r}_{*i}, \mathbf{P}) \| p(\mathbf{z}_i))]
\end{aligned}
\label{ob_item}
\end{equation}
where $p_{\theta_i}(\mathbf{r}_{*i} | \mathbf{z}_{1:|\mathcal{U}|}, \mathbf{z}_i, \mathbf{P}) = \prod_u p_{\theta_i}(r_{u,i}| \mathbf{z}_u, \mathbf{z}_i, \mathbf{p}_i)$. Note the above objectives $\mathcal{L}^u_{vae}$ and $\mathcal{L}^i_{vae}$ are not conflicting with each other, since maximizing either of them corresponds to the maximization of the overall ELBO. 

\subsection{Neighborhood-enhanced Representation Constraint}
The NRC module is designed to prevent the confusion between latent representations inferred from different aspects and maintain correspondence between aspect-level representations of users and items. Specifically, we first aggregate the interacted neighbors to calculate the aspect-level neighborhood-based representations $\mathbf{o}^a_u$ and $\mathbf{o}^a_i$ of user $u$ and item $i$,
\begin{equation}
\mathbf{o}^a_u = \sum\nolimits_{i \in \mathcal{N}_u} c^a_i \cdot \mathbf{z}^a_i, ~~~ \mathbf{o}^a_i = \sum\nolimits_{u \in \mathcal{N}_i} p^a_u \cdot \mathbf{z}^a_u,
\end{equation}
where $\mathcal{N}_u$ and $\mathcal{N}_i$ are the neighbor sets of user $u$ and item $i$, respectively. Afterwards, we treat the inferred aspect-level latent representations $\mathbf{z}^a_u$ and neighborhood-based representation $\mathbf{o}^a_u$ of user $u$ as positive sample pair. Moreover, we set two kinds of negative samples. Firstly, to constrain the independence among different aspect representations, we set aspect-level negative samples by taking representations of different aspects for the same user as negative samples. Secondly, we use representations from different users under the same aspect as user-level negative samples to guarantee the discrepancy in aspect-level representations among different users. The InfoNCE~\cite{gutmann2010noise} loss is employed to achieve the contrastive constraint,
\begin{equation}
	\mathcal{L}^a_u = -\log \frac{e^{\frac{s(\mathbf{z}^a_u,\mathbf{o}^a_u)}{\tau}}}
	{\underbrace{e^{\frac{s(\mathbf{z}^a_u,\mathbf{o}^a_u)}{\tau}}}_{\text{pos-pair}} + 
	\underbrace{\sum\limits_{b \neq a}^{A} e^{\frac{s(\mathbf{z}^a_u,\mathbf{o}^b_u)}{\tau}}}_{\text{aspect-level neg-pairs}}
	+ \underbrace{\sum\limits_{v \neq u}^{B_u} e^{\frac{s(\mathbf{z}^a_u,\mathbf{o}^a_v)}{\tau}}}_{\text{user-level neg-pairs}}}
  \label{cl_user}
\end{equation}
where $\mathcal{L}^a_u$ is the contrastive loss of each aspect-level representation of user $u$, $s(\cdot)$ denotes the cosine similarity measure, $\tau$ is a temperature parameter (generally set to $0.2$), and $B_u$ denotes a set of users within a batch. Similarly, the item disentangled representation can be constrained by,
\begin{equation}
	\mathcal{L}^a_i = -\log \frac{e^{\frac{s(\mathbf{z}^a_i,\mathbf{o}^a_i)}{\tau}}}
	{\underbrace{e^{\frac{s(\mathbf{z}^a_i,\mathbf{o}^a_i)}{\tau}}}_{\text{pos-pair}} + 
	\underbrace{\sum\limits_{b \neq a}^{A} e^{\frac{s(\mathbf{z}^a_i, \mathbf{o}^b_i)}{ \tau}}}_{\text{aspect-level neg-pairs}}
	+ \underbrace{\sum\limits_{j \neq i}^{B_i} e^{\frac{s(\mathbf{z}^a_i, \mathbf{o}^a_j)}{\tau}}}_{\text{item-level neg-pairs}}}
\label{cl_item}
\end{equation}
where $B_i$ denotes a set of items within a batch. Intuitively, the auxiliary supervision of positive pairs encourages the direction consistency between latent representations of users and items on the same aspect, while the supervision of negative pairs enforces
the divergence among different aspects. Note there are also some methods~\cite{wang2020disentangled} that use distance correlation as a regularizer to achieve representation independence modeling. We do not use such distance correlation, because it can only ensure the independence of the aspects, but cannot guarantee the correspondence between the aspect-level representations of users and items. 

Finally, the overall objects for alternately optimization can be set as,
\begin{equation}
	\mathcal{L}_u = \mathcal{L}^u_{vae} + \gamma \cdot \sum\nolimits_{a=1}^{A} \mathcal{L}^a_u, ~~~ \mathcal{L}_i = \mathcal{L}^i_{vae} + \gamma \cdot \sum\nolimits_{a=1}^{A} \mathcal{L}^a_i,
\end{equation}
where $\gamma$ is an adjusted factor to balance the VAE loss and the contrastive loss.

	\section{Experiment}
	\label{sec:experiment}
	% !TEX spellcheck = en_US
% !TeX root = ../main.tex

\begin{table}[t]
	\small
	\centering
	\caption{Statistics of three evaluation datasets.}
	\setlength{\tabcolsep}{0.8mm}{
		\begin{tabular}{lrrrr}
			\hline
			\hline
			\textbf{Dataset} & \textbf{\#Users} & \textbf{\#Items} & \textbf{\#Feedback} & \textbf{Sparsity}\\
			\hline
			\textbf{ML1M}    & 6,040  & 3,679  & 1,000,180  & 0.9550  \\
			\textbf{AKindle}  & 14,356 & 15,885 & 367,477   & 0.9984  \\
			\textbf{Yelp} & 31,668 & 38,048 & 1,561,406  & 0.9987  \\
			\hline
			\hline
	\end{tabular}}
	\label{tab:dataset}
\end{table}

\setlength{\tabcolsep}{0.08mm}
\begin{table*}[t]
	\small
	\centering
	\caption{Performance comparisons of DualVAE \emph{vs.} baselines. Best performance is in boldface. Improvement is obtained between DualVAE and the best result (underlined) in baselines. $*$ indicates that the improvement is significant with $p<0.05$.}
	\begin{tabular}{l|cccc|cccc|cccc}
		\hline
		\hline
		\textbf{Datasets}  & \multicolumn{4}{c|}{\textbf{ML1M}} & \multicolumn{4}{c|}{\textbf{AKindle}} & \multicolumn{4}{c}{\textbf{Yelp}} \\
		\hline
		\textbf{Methods} & \textbf{R@20} & \textbf{N@20} & \textbf{R@50} & \textbf{N@50} & \textbf{R@20} & \textbf{N@20} & \textbf{R@50} & \textbf{N@50} & \textbf{R@20} & \textbf{N@20} & \textbf{R@50} & \textbf{N@50} \\
		\hline
		\textbf{MF}~\cite{koren2009matrix} & 0.1846  & 0.3122  & 0.3193  & 0.3217  & 0.0453  & 0.0364  & 0.0707  & 0.0398  & 0.0298  & 0.0256  & 0.0589  & 0.0362  \\
		\textbf{NeuMF}~\cite{he2017neural} & 0.2158  & 0.3256  & 0.3569  & 0.3472  & 0.0676  & 0.0459  & 0.1073  & 0.0582  & 0.0547  & 0.0456  & 0.1060  & 0.0622  \\
		\textbf{JCA}~\cite{zhu2019improving} & 0.2251  & 0.3303  & 0.3814  & 0.3612  & 0.0745  & 0.0464  & 0.1171  & 0.0626  & 0.0550  & 0.0452  & 0.1121  & 0.0663  \\
		\hline
		\textbf{PoissVAE}~\cite{liang2018variational} & 0.2286  & 0.3439  & 0.3790  & 0.3622  & 0.0623  & 0.0392  & 0.1084  & 0.0532  & 0.0556  & 0.0453  & 0.1076  & 0.0648  \\
		\textbf{MultiVAE}~\cite{liang2018variational} & 0.2301  & 0.3348  & 0.3819  & 0.3575  & 0.0739  & 0.0459  & 0.1226  & 0.0607  & 0.0563  & 0.0455  & 0.1091  & 0.0653  \\
		\textbf{MacridVAE}~\cite{ma2019learning} & \underline{0.2313}  & 0.3409  & \underline{0.3865}  & 0.3655  & \underline{0.0779}  & 0.0475  & \underline{0.1335}  & \underline{0.0654}  & \underline{0.0601}  & \underline{0.0485}  & \underline{0.1165}  & \underline{0.0681}  \\
		\textbf{JoVA}~\cite{askari2021variational} & 0.2305  & 0.3409  & 0.3840  & 0.3641  & 0.0754  & 0.0470  & 0.1270  & 0.0631  & 0.0573  & 0.0459  & 0.1122  & 0.0665  \\
		\textbf{BiVAE}~\cite{truong2021bilateral} & 0.2305  & \underline{0.3450}  & 0.3853  & \underline{0.3658}  & 0.0757  & \underline{0.0477}  & 0.1295  & 0.0636  & 0.0575  & 0.0460  & 0.1140  & 0.0670  \\
		\hline
		\textbf{DualVAE} & \textbf{0.2365*} & \textbf{0.3643*} & \textbf{0.3944*} & \textbf{0.3816*} & \textbf{0.0812*}  & \textbf{0.0511*}  & \textbf{0.1378*} & \textbf{0.0683*}  & \textbf{0.0633*} & \textbf{0.0514*}  & \textbf{0.1227*} & \textbf{0.0735*} \\
		\hline
		\textbf{Improvement} & 2.27\% & 5.59\% & 2.04\% & 4.32\% & 4.24\% & 7.13\% & 3.21\% & 4.37\% & 5.32\% & 5.98\% & 5.32\% & 7.93\% \\
		\hline
		\hline
	\end{tabular}%
	\label{tab:compared}
\end{table*}%

\subsection{Experimental Setup}
\subsubsection{\textbf{Datasets and Metrics.}}
The experimental study of DualVAE is conducted on three publicly available benchmark datasets from different platforms: MovieLens-1M (\textbf{ML1M} for short), Amazon Kindle Store (\textbf{AKindle} for short) and \textbf{Yelp}. The first dataset is a MovieLens\footnote{https://grouplens.org/datasets/movielens} dataset, where each user has at least 20 interactions, the second dataset is collected from Amazon\footnote{http://jmcauley.ucsd.edu/data/amazon}~\cite{ni2019justifying}, and the third dataset is from the 2018 Yelp challenge\footnote{https://www.yelp.com/dataset.}. For the last two datasets, we use a 10-core setting to ensure their data quality.
For all datasets, we treat observed user-item interactions as positive feedback. Table~\ref{tab:dataset} summarizes the statistics of the three evaluation datasets. The performance of all models on the testing set is evaluated by two commonly used metrics: Recall (R@\textit{N}) and Normalized Discounted Cumulative Gain (N@\textit{N})~\cite{he2015trirank}. We truncate the ranked list by setting $N$ at $\{20, 50\}$. The learned recommendation model can get a ranked top-\textit{N} list from all items to evaluate the two metrics.

\subsubsection{\textbf{Baselines.}}
We compare DualVAE versus the following two groups of competitive methods: 1) Latent factor model-based CF methods, including \textbf{MF}~\cite{koren2009matrix}, \textbf{NeuMF}~\cite{he2017neural} and \textbf{JCA}~\cite{zhu2019improving}; 2) VAE-based CF methods, including \textbf{PoissVAE} and \textbf{MultiVAE}~\cite{liang2018variational}, \textbf{MacridVAE}~\cite{ma2019learning}, \textbf{JoVA}~\cite{askari2021variational}, and \textbf{BiVAE}~\cite{truong2021bilateral}. Note for a fair comparison, PoissVAE, BiVAE and DualVAE all use Poisson likelihood by default in their generative models. For all the baselines, we use the implementations and parameter settings reported in their original papers.

\subsubsection{\textbf{Parameter Settings.}}
Our model is implemented in Pytorch\footnote{https://github.com/georgeguo-cn/DualVAE}. For all datasets, we randomly select 80\% user interactions for training and the remaining for testing. From the testing set, we randomly select $10\%$ interactions as validation set to tune hyperparameters. For a fair comparison, the total embedding size is fixed to $A \times d=100$ and the mini-batch Adam~\cite{kingma2014adam} is employed to update model parameters with a fixed batch size of $128$ for all models. The learning rate is searched from $\{1e^{-4}, 1e^{-3}, \dots, 1e^{-1}\}$. We search the coefficient $\gamma$ from $\{1e^{-5}, 1e^{-4}, \dots, 1e^{-1}\}$ and search the aspect number $A$ from $\{1, 2, 4, 5, 10, 20\}$. We perform the grid search of hyperparameters to obtain the optimal set on different datasets. In addition, all experiments in this paper are performed in the same experimental environment with Intel(R) Xeon(R) Silver 4210R CPU @ 2.40GHz and GeForce RTX 3090.
 
\subsection{Performance Comparison}
Table~\ref{tab:compared} presents the overall performance comparison, from which we have the following key observations:
\begin{itemize}[leftmargin=1em]
	\item \textit{DualVAE consistently outperforms all the other baselines on all datasets and metrics}. In particular, DualVAE's improvement over the best baseline is statistically significant in all cases, demonstrating that disentangling multi-aspect features of users and items indeed can help improve performance. Moreover, as the sparsity of datasets increases, the improvement of DualVAE becomes more significant, indicating that capturing multi-aspect uncertainty of user and item feature spaces can further alleviate the data sparsity problem and improve model robustness. 
	
	\item \textit{Dual VAE-based methods outperform user-specific VAE-based methods on all datasets}. Specifically, the performance of BiVAE is higher than that of PoissVAE and MultiVAE, while DualVAE achieves better performance than MacridVAE. The results indicate that inferring latent representations of both users and items can better adapt to the two-way nature of interaction data, thereby promoting accuracy.  
	
	\item \textit{Disentanglement-based methods generally achieve better performance than their non-disentanglement counterparts}. For instance, MacridVAE outperforms MultiVAE by an average of 3.97\% in terms of N@$20$ across all datasets, indicating that disentangling multi-aspect user preferences is beneficial for improving representation quality and performance.
\end{itemize}

\setlength{\tabcolsep}{0.5mm}
\begin{table}[t]
	\small
	\centering
	\caption{Performance of various variants of DualVAE.}
	\begin{tabular}{l|c|c|c|c|c|c}
		\hline
		\hline
		\textbf{Datasets} & \multicolumn{2}{c|}{\textbf{ML1M}} & \multicolumn{2}{c|}{\textbf{AKindle}} & \multicolumn{2}{c}{\textbf{Yelp}} \\
		\hline
		\textbf{Methods} & \textbf{R@20} & \textbf{N@20} & \textbf{R@20} & \textbf{N@20} & \textbf{R@20} & \textbf{N@20} \\
		\hline
		\textbf{w/o ADD} & 0.2305  & 0.3450  & 0.0757  & 0.0473  & 0.0575  & 0.0460 \\
		\textbf{w/o ID} & 0.2325  & 0.3594  & 0.0801  & 0.0495  & 0.0602  & 0.0489  \\
		\textbf{w/o UD} & 0.2316  & 0.3492  & 0.0772  & 0.0483  & 0.0574  & 0.0463  \\
		\hline
		\textbf{w/o NRC} & 0.2321  & 0.3601  & 0.0803  & 0.0501  & 0.0611  & 0.0493  \\
		\textbf{w/o UNS} & 0.2354  & 0.3632  & 0.0805  & 0.0503  & 0.0619  & 0.0502  \\
		\textbf{w/o ANS} & 0.2353  & 0.3632  & 0.0810  & 0.0509  & 0.0624  & 0.0510  \\
		\textbf{w/o NPS} & 0.2345 & 0.3623 & 0.0808 & 0.0508 & 0.0620 & 0.0503 \\
		\hline
		\textbf{DualVAE} & \textbf{0.2365} & \textbf{0.3643} & \textbf{0.0812} & \textbf{0.0511} & \textbf{0.0633} & \textbf{0.0514} \\
		\hline
		\hline
	\end{tabular}%
	\label{tab:ablation}%
\end{table}%

\subsection{Ablation Studies}
Table~\ref{tab:ablation} reports the results of ablation studies for various variants of DualVAE. Specifically, \textbf{w/o ID} (\textbf{w/o UD}) denotes that our model only retains user(item)-side aspect disentanglement; \textbf{w/o UNS} and \textbf{w/o ANS} are two variants that do not use user-level and aspect-level negative samples in NRC module, respectively. \textbf{w/o NPS} is a variant that removes neighborhood-based representations and uses inferred aspect-level representations and themselves to construct positive pairs. From Table~\ref{tab:ablation}, we can find:
\begin{itemize}[leftmargin=1em]
 \item After removing the ADD module, the performance of DualVAE drops (\emph{vs.} \textbf{w/o ADD}) by an average of 6.65\% and 8.46\% respectively on R@$20$ and N@$20$, indicating the effectiveness of attention-aware disentanglement. The performance gap between DualVAE and \textbf{w/o UD}  (\textbf{w/o ID}) demonstrates the benefit of shaping the aspect distributions of both users and items in improving performance. Moreover, \textbf{w/o ID} performs better than \textbf{w/o UD} on sparser datasets, which suggests that the user-side disentanglement may bring more performance gains than the item-side to alleviate interaction sparsity.
 
 \item The variant \textbf{w/o NRC} removes the NRC module and causes an average performance drop of 2.21\% and 2.47\% in terms of R@$20$ and N@$20$ respectively, which indicates the necessity of ensuring the correspondence and independence of disentangled representations in promoting performance. The comparison between DualVAE and \textbf{w/o ANS} (\textbf{w/o UNS}) reveals that it is beneficial to maintain the independence of disentangled representations by exploiting two-level negative samples. Moreover, \textbf{w/o ANS} performs slightly better than \textbf{w/o UNS} on AKindle and Yelp, showing that aspect-level differences may be more important than user-level differences in keeping the representation independence. \textbf{w/o NPS} performs worse than DualVAE, which demonstrates the significance of maintaining the correspondence between multi-aspect representations of users and items.
\end{itemize}

\begin{figure}[t]
	\centering
	\subfloat[\scriptsize{Impact of aspect number $A$.}]{
		\includegraphics[width=0.24\textwidth]{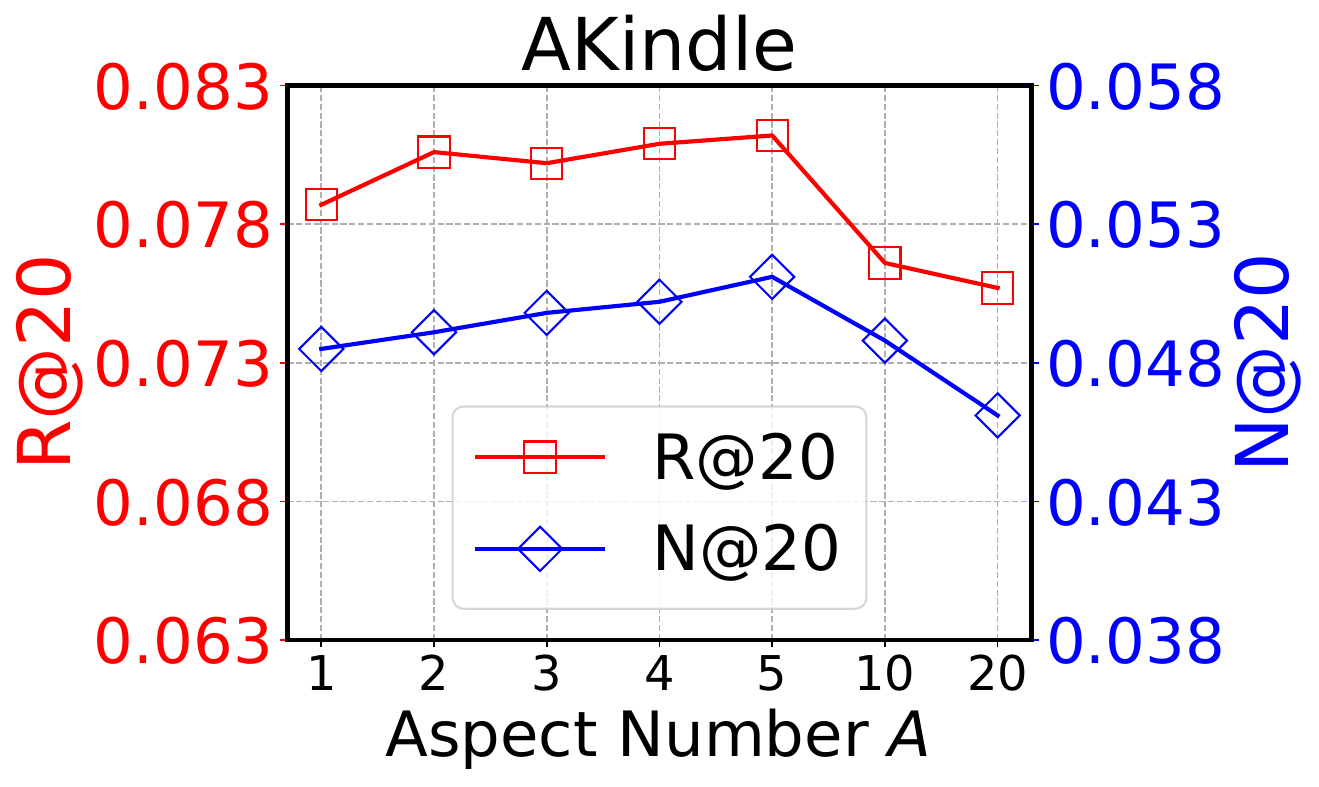}
		\includegraphics[width=0.24\textwidth]{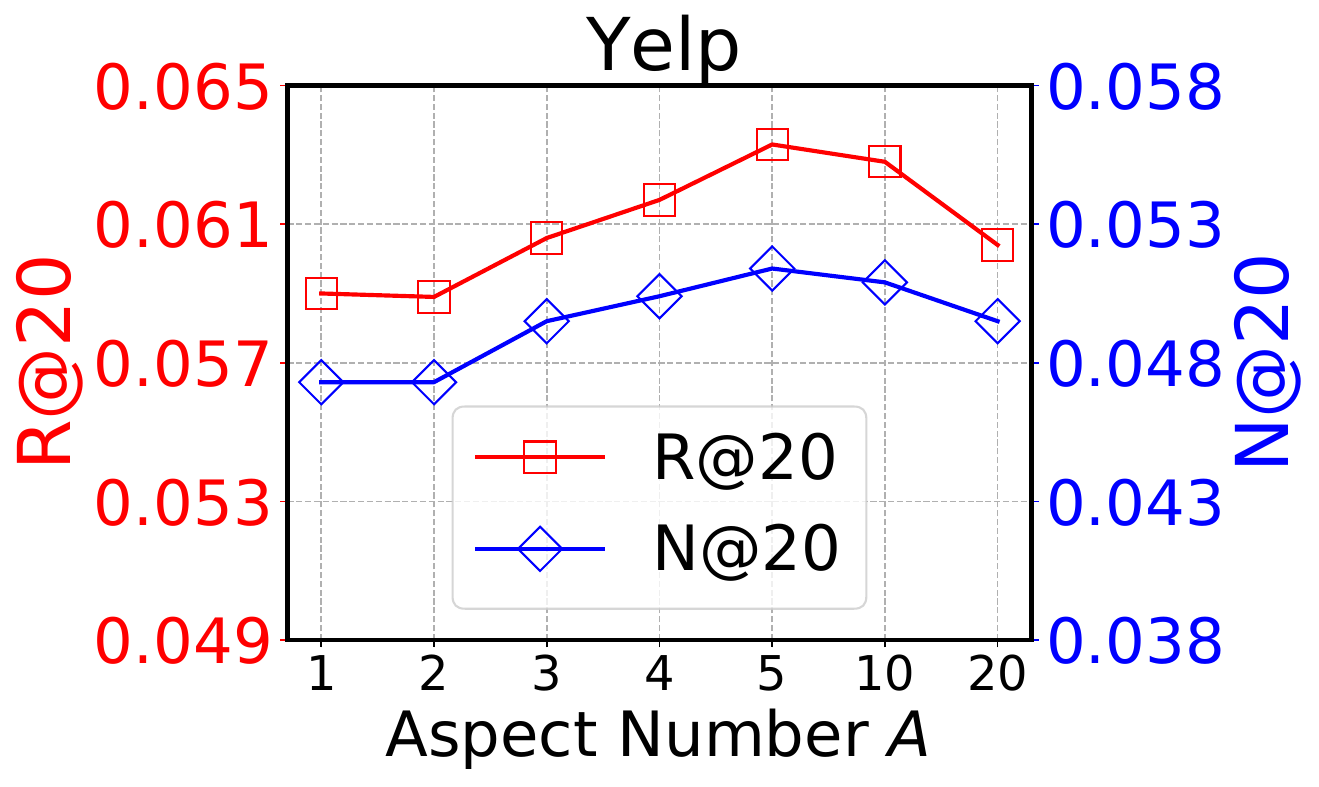}
	}\\
	\subfloat[\scriptsize{Impact of balance coefficient $\gamma$}]{
		\includegraphics[width=0.24\textwidth]{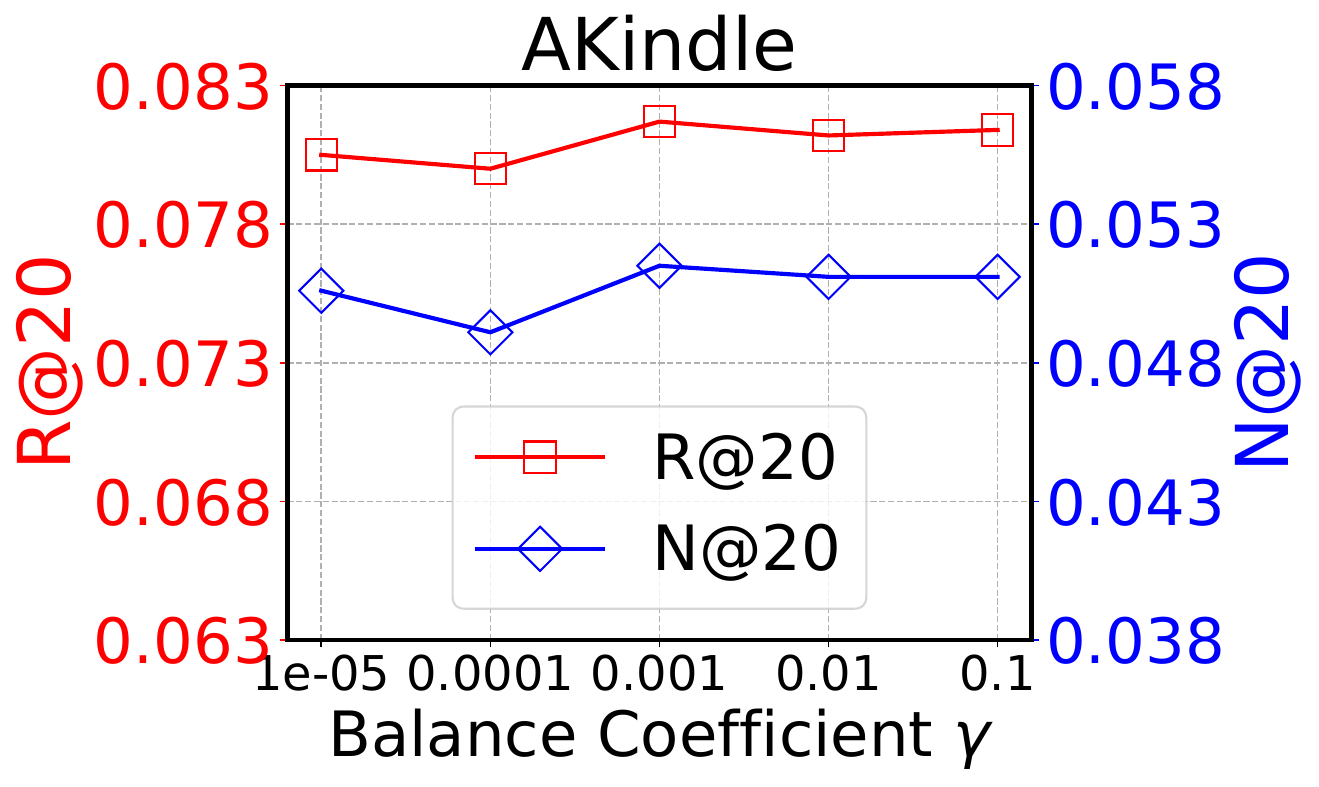}
		\includegraphics[width=0.24\textwidth]{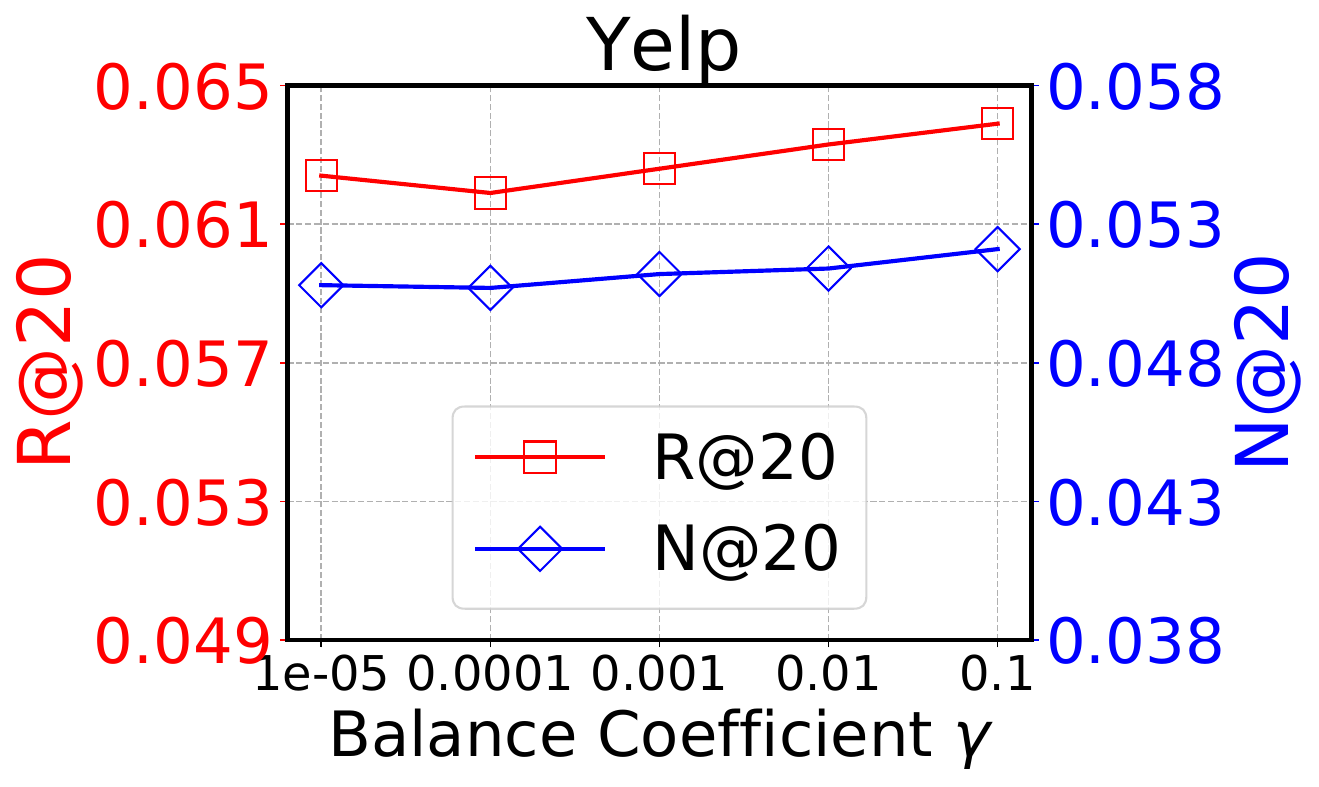}
	}
	\caption{Performance of varying parameters $A$ and $\gamma$ in terms of R@$20$ and N@$20$ on AKindle and Yelp.}
	\label{fig:paramer}
\end{figure}

\subsection{Hyperparameter Analysis}
\subsubsection{\textbf{Impact of Aspect Number $A$.}}
To investigate the effect of aspect number, we vary $A$ in the range $\{1, 2, 3, 4, 5, 10, 20\}$ and show the performance comparisons in Figure~\ref{fig:paramer}~(a). We can observe that the poor performance is achieved when $A=1$ on the datasets Akindle and Yelp, which indicates that one aspect generally is insufficient for fitting interaction data. Overall, the performance first increases with the growth of the number of aspects, and then declines after reaching the optimum ($A=5$). The results prove that disentangled representation learning is beneficial, but too many aspects may be redundant and will negatively affect performance. The results on ML1M show a similar trend, and the optimal performance is obtained at $A=10$.

\subsubsection{\textbf{Impact of Balance Coefficient $\gamma$.}}
The parameter $\gamma$ is an adjusted coefficient of contrastive loss. To evaluate the effect of $\gamma$, we search for it from $\{1e^{-5}, 1e^{-4}, \dots, 1e^{-1}\}$. Figure~\ref{fig:paramer}~(b) presents the results in terms of R@$20$ and N@$20$ on the datasets Akindle and Yelp. From the results, we can see that a larger $\gamma$ with a stronger independence constraint is desirable for reaching a better performance. In particular, we can set $\gamma=0.1$ for ML1M and Yelp, and $\gamma=0.001$ for AKindle to learn a suitable model.

\begin{figure}[t]
	\centering
	\includegraphics[width=0.48\textwidth]{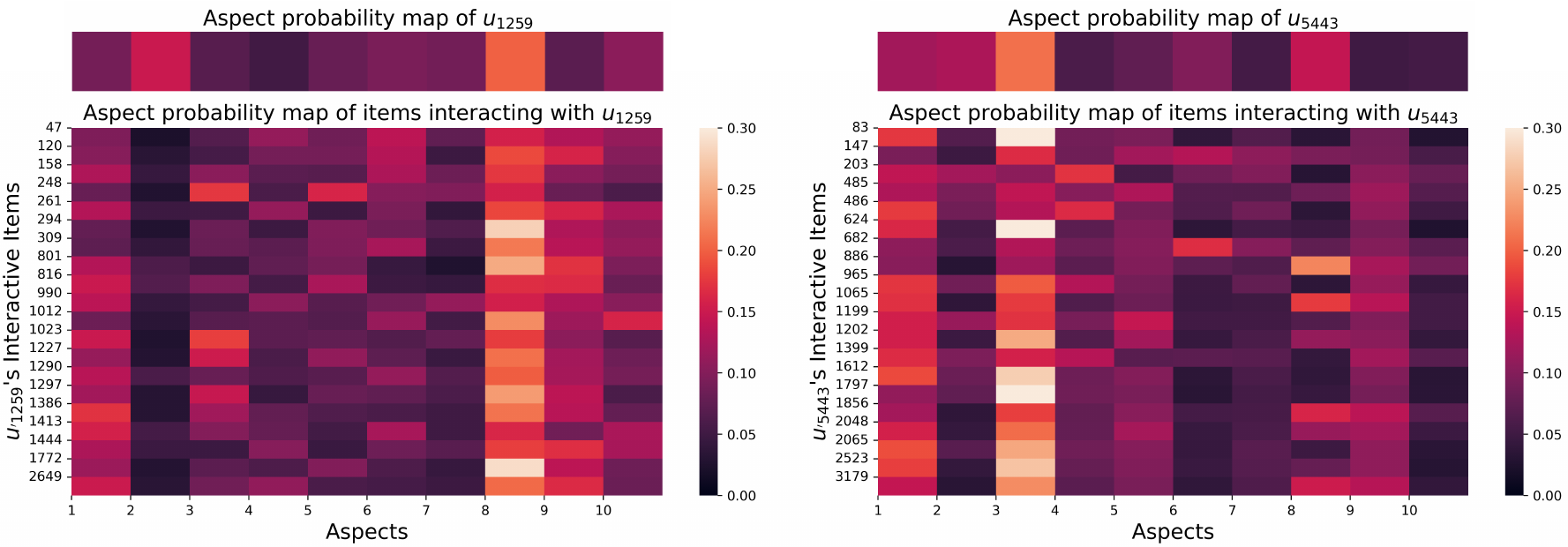}
	\caption{Visualization of aspect probability maps of $u_{1259}$, $u_{5443}$ and their random $20$ interactive items on ML1M dataset.}
	\label{fig:probability}
\end{figure}

\begin{figure}[t]
	\centering
	\includegraphics[width=0.48\textwidth]{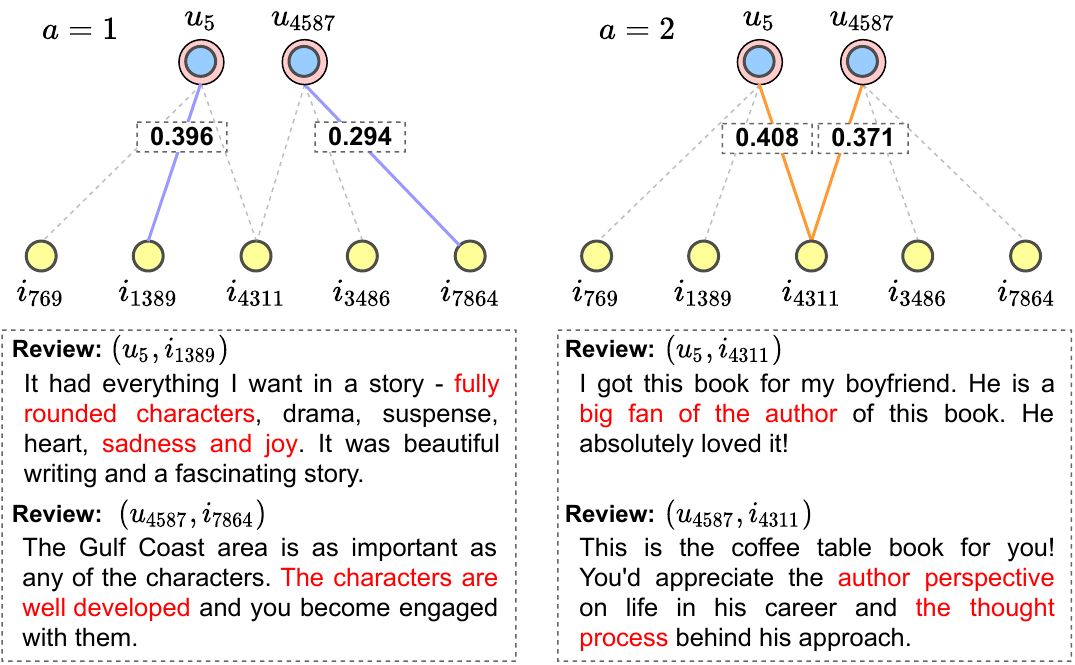}
	\caption{Case studies of our proposed DualVAE on AKindle dataset. Red words reflect users' attitude in their reviews towards certain aspects of an item.}
	\label{fig:case}
\end{figure}

\subsection{Interpretability Exploration} 
\subsubsection{Visualization of Aspect Probability.}
We first visualize the aspect probability maps of two specific users  $u_{1259}$ and $u_{5443}$ in ML1M, as well as their random $20$ interactive items, as illustrated in Figure~\ref{fig:probability}. We can see that user $u_{1259}$ pays more attention to the  $8$-th aspect, while $u_{5443}$ focuses more on the $3$-rd aspect, which demonstrates that our model indeed can learn personalized user preferences. Furthermore, the aspect probability distributions are similar between users and most of the items they interacted with. In particular, most of $u_{1259}$'s interactive items have a higher probability (brighter in the map) in the $8$-th aspect, while $u_{5443}$ and most of its interactive items have a better match in the $3$-rd aspect. This phenomenon suggests that some specific aspects may dominate the interactions between users and items, which provides a good perspective for explaining user behaviors.

\subsubsection{Interpretability of Disentangled Representations.}
We further explore the interpretability of disentangled representations by a case study, as shown in Figure~\ref{fig:case}. Specifically, we take the AKindle dataset as example, and randomly select two users, $u_5$ and $u_{4587}$ and their interactive items from it to explain their behaviors by analyzing their reviews. Specifically, we present the corresponding user reviews for the interactive items with the highest aspect-level prediction score under two random aspects. For example, interaction $(u_5, i_{1389})$ has the highest predicted score (thick solid line with color) under the $1$-st aspect ($a = 1$), which implies that the occurrence of this interaction is more related to the matching of user's preference and item's characteristic under the $1$-st aspect. From the results, we have the following findings:
\begin{itemize}[leftmargin=1em]
    \item By analyzing different reviews under the same aspect, we can find that the reviews seem to be consistent with some intuitive attribution concepts. Specifically, we can summarize the semantics of the two implicit aspects from the reviews (especially the red words) as: \textit{character} and \textit{author}. The results interpret the capability of DualVAE to disentangle multi-aspect preferences from user behaviors.
    
    \item A higher prediction score in general corresponds to a higher degree of aspect-level feature matching between users and items. For example, under the $2$-nd aspect, both $u_{5}$ and $u_{4587}$ really appreciate the author of item $i_{4311}$, which is a good post-hoc explanation for their purchasing of item $i_{4311}$. The results further verify the rationality of multi-aspect feature matching between users and items.
\end{itemize}

	\section{Conclusion}
	\label{sec:conclusion}
	% !TEX spellcheck = en_US
% !TeX root = ../main.tex

In this work, we proposed DualVAE, which combines disentangled representation learning with VAE to fit user-item interactions. Specifically, we first designed an attention-aware dual disentanglement module and unified it with a disentangled variational autoencoder to infer multi-aspect latent representations of both users and items for reconstructing the observed interactions. Moreover, we developed a neighborhood-enhanced representation constraint module to ensure the quality of disentangled representations by a contrastive learning with a neighborhood-based positive sample and two-level negative samples. Extensive experiments on three real-world datasets demonstrate the effectiveness of DualVAE. Further empirical studies also explore the interpretability of disentangled representations. For future work, we plan to introduce more content knowledge, such as user reviews, modality information, to establish ground-truth features of users and items for more fine-grained disentangled representations.
	
	\section{Acknowledgements}
	We would like to thank all anonymous reviewers for their valuable comments. The work was partially supported by the National Natural Science Foundation of China under Grant No. 62272176 and the National Key R\&D Program of China under Grant No. 2022YFC3802101 and 2023YFB3308301.
	
	\bibliographystyle{siam}
	\bibliography{myref}

%	\clearpage
%	\input{section/appendix}
		
\end{document}